\documentclass[twocolumn,showpacs,amsmath,amssymb,aps]{revtex4}

\usepackage{graphicx}
\usepackage{dcolumn}
\usepackage{bm}

\newcommand{\rl}{\leftrightarrow}

\newcommand\mean[1]{{\langle#1\rangle}}

\begin{document}


\title{Emergence of pulled fronts in fermionic microscopic particle models}

\author{Esteban Moro} \email{emoro@math.uc3m.es}
\affiliation{%
Grupo Interdisciplinar de Sistemas Complejos (GISC) and  Departamento de Matem\'aticas, Universidad Carlos III de
  Madrid, Avda.\ de la Universidad 30, E-28911 Legan\'es, Spain
}%

\date{\today}

\begin{abstract}
  We study the emergence and dynamics of pulled fronts described by
  the Fisher-Kolmogorov-Petrovsky-Piscounov (FKPP) equation in the
  microscopic reaction-diffusion process $A + A \rl A$ on the lattice
  when only a particle is allowed per site. To this end we identify
  the parameter that controls the strength of internal fluctuations in
  this model, namely, the number of particles per correlated volume.
  When internal fluctuations are suppressed, we explictly see the
  matching between the deterministic FKPP description and the
  microscopic particle model.
\end{abstract}


\maketitle

The study of diffusion-limited reaction processes has shown the
important role of internal or microscopic fluctuations in low
dimensions \cite{marro,avraham,hinrichsen}. Mean-field approximation
for those processes assume that diffusive mixing is much stronger than
the influence of correlations produced by reactions.  However,
diffusive mixing is not strong enough in low-dimensional systems and
fluctuations might modify the dynamics or induce nonequilibrium phase
transitions. While this behavior is observed in different situations,
there is an special interest in the problem of front propagation in
reaction-diffusion systems
\cite{avraham,doering1,riordan,moro,saarloos,brunet,breuer,kessler,avraham1,panja,tripathy,bramson86,kerstein,kerstein88,lemarchand}.
In this paper we concentrate in microscopic lattice reaction-diffusion
models whose mean-field approximation is given by the FKPP equation
\cite{fisher}
\begin{equation}\label{FKPPeq}
\frac{\partial \rho}{\partial t} = D \frac{\partial^2 \rho}{\partial x^2} + k_1 \rho - k_2 \rho^2,
\end{equation}
where $\rho(x,t)$ is the local concentration of particles. Such an
equation display traveling-wave solutions of the form $\rho =
\rho(\xi)$ with $\xi = x-vt$ which invade the unstable phase $\rho =0$
from the stable phase $\rho = k_1/k_2$ and travel with velocity $v\geq
v_0 = 2\sqrt{D k_1}$. For steep enough initial conditions, the
solution selected for large times is the one with minimal velocity,
$v_0$, which is known to be a pulled front, since it is essentially
``pulled along'' by the growth and spreading of small perturbations in
the leading edge where $\rho \ll 1$ \cite{ebert,saarloos}.
Microscopic fluctuations are expected to modify macroscopic properties
of pulled fronts at two levels: (i) first because the deterministic
description (\ref{FKPPeq}) breaks down at small densities $\rho \sim
1/N$ where $N$ is the number of particles, which introduces an
effective cutoff in the FKPP equation. Due to the importance of the
tail development in pulled fronts, several front features are
dramatically affected by this effective cutoff \cite{brunet}: for
example, the selected velocity converges as $\ln^{-2} N$ to the
mean-field value $v_0$, (ii) second, because internal fluctuations
are present and could interplay with or even destroy pulled front
development and dynamics \cite{doering1,avraham1,moro}.


One of the most studied microscopic models is the reversible reaction
model $A \rl A + A$ on a lattice
\cite{doering1,riordan,avraham1,breuer,kessler,panja,bramson86,kerstein,kerstein88,lemarchand,tripathy,moro}.
In the bosonic version of this model \cite{breuer}, the number of
possible particles per site is unbounded and thus the balance between
birth and coagulation gives an average number of particles per site
$N$. If $N \to \infty$ the reaction is well stirred within each site
and the front dynamics is described by the mean-field approximation
(\ref{FKPPeq}). For very large $N$ discreteness effects remain and
produce the predicted velocity correction $v - v_0 \sim \ln^{-2} N$
\cite{brunet,kessler}. In the case of the fermionic version of the $A
\rl A+A$ model only a particle is allowed per site.  The main reason
to consider exclusion is that, for some values of the parameters, the
model is analytically tractable
\cite{avraham,doering1,avraham1,bramson86} and/or simulations are
easier than in the bosonic version. Exact results are available for
the two interesting regimes in the model: reaction-limited regime
\cite{bramson86}, where coarse-grained density front profiles are
described by the mean-field FKPP equation, and diffusion-limited
regime \cite{avraham,doering1,avraham1} in which internal fluctuations
dominate front propagation and the mean-field approximation
(\ref{FKPPeq}) is not valid.  Our purpose in this paper is to put
these two results in a general framework that can describe the
emergence of pulled fronts in this fermionic model. This is done by
identifying the control parameter that modulates the effect of
internal fluctuations on the front propagation model. As we will see,
this parameter also controls the development of the tail front and
establishes the appearance of pulled fronts.

In the $A \rl A + A$ model in one dimension, particles are allowed to
occupy lattice sites and can undergo the following moves: (i)
Diffusion to any one of its two neighbor lattice sites with a
diffusion rate $D$ (ii) Birth: occupied sites spontaneously generate
particles at neighbor lattice sites with rate $\mu$, (ii) Coagulation:
a particle can get annihilated with death rate $\eta$ if one of its
two neighboring filled lattice sites is occupied. The fermionic nature
of the model makes diffusion and birth only possible if the
neighboring lattice site is empty. The mean-field description of this
model is given by the FKPP equation with $k_1 = 2 \mu$, $k_2 = 2 (\mu
+ \eta)$. Starting from an initial condition in which occupation
number is only different fron zero on the right side of a site, a
front develops and advances as a function on time.
Operationally, the front position $x_f(t)$ is determined by a local
average of density of particles over intervals of length $\lambda^{-1}
= 2D/v_0$, which is the width of the deterministic front selected for
our initial condition by the FKPP equation (\ref{FKPPeq})
\cite{ebert}. Thus, front position is the point where this
coarse-grained density equals $\rho_0/2$ \cite{tripathy,moro}. Other
definitions of the front position yield the same results
\cite{bramson86}. Since results only depend on the ratios $D/\mu$ and
$D/\eta$ we set $D=1/2$ throughout this paper. After a transient time
(which could be long), the front advances linearly, i.e.\ 
$\mean{x_f(t)} = v t$, where $\mean{\cdots}$ stands for average over
different realizations. In this regime, statistical properties of the
front with respect to the normalized coordinate $\xi = x - v t$, are
independent of time.

\begin{figure}
\begin{center}
\includegraphics[width=3.0in,clip=]{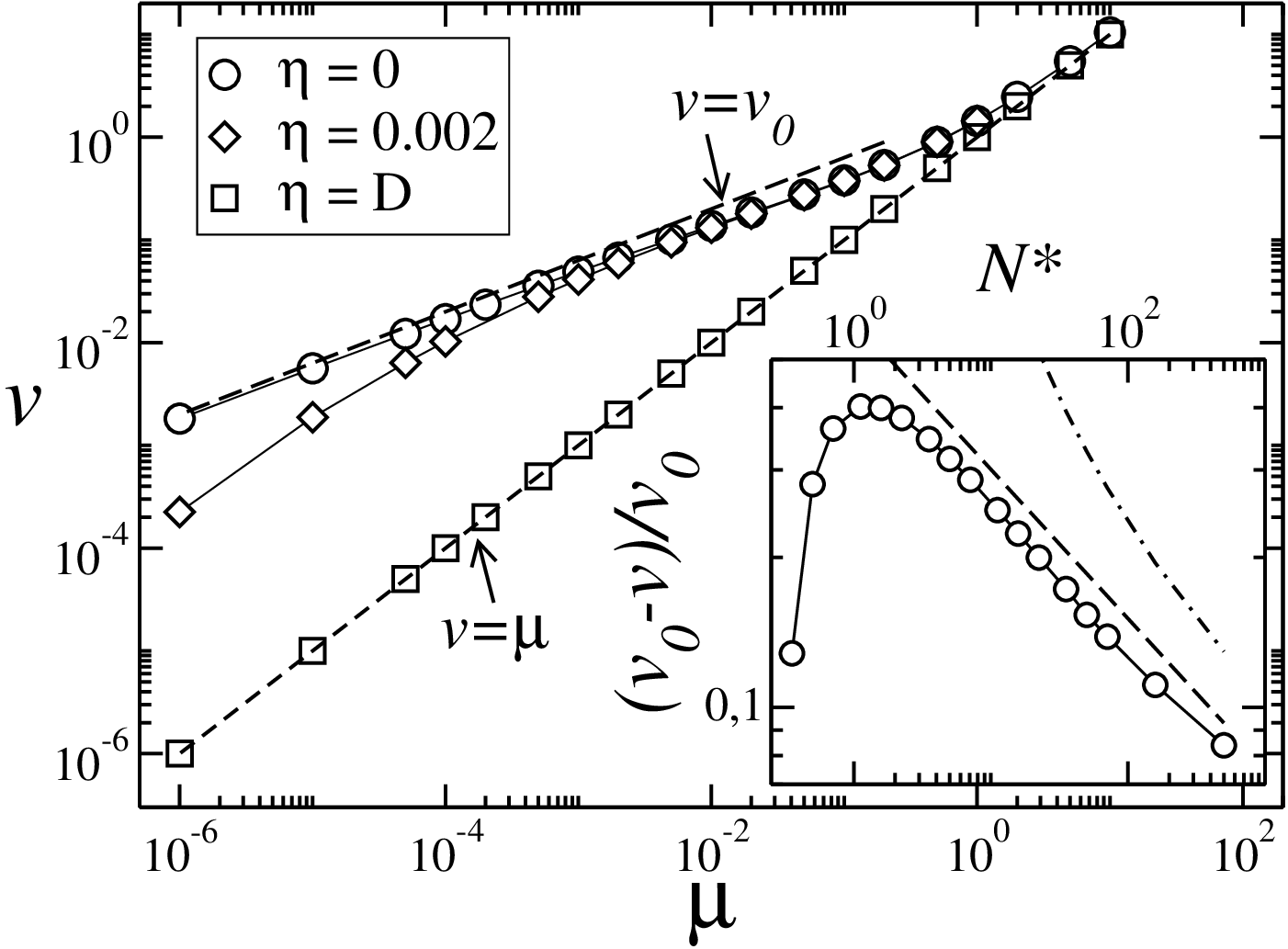}
\caption{\label{vel}
  Velocity as a function of the birth rate $\mu$ for different values
  of the coagulation rate $\eta$. Simulations are done with $D = 0.5$.
  Dashed lines correspond to the predictions $v = \mu$
  \cite{doering1,avraham} and $v = v_0$ \cite{bramson86}.
  Inset: Velocity corrections as a function of $\mu$ for the case
  $\eta = 0$. Dashed line is the power law $N^{-1/3}$ while
  dash-dotted line is the prediction $\pi^2/(2\ln^{2}N^{*})$ of
  \cite{brunet}.}
\end{center}
\end{figure}

In this model, there are two special cases for which exact results are
available: when $\eta = D$ the model is solvable using the method of
inter-particle distribution functions \cite{doering1,avraham}. In that
case fronts advance with velocity $v = \mu $ which shows how internal
fluctuations can dominate the system behavior in this
diffusion-limited regime. On the other hand, when $\eta = 0$ it was
proved in \cite{bramson86} that fronts approach asymptotically the
FKPP equation predictions ($v=v_0$) in the limit $D/\mu \to \infty$
(reaction-limited regime), while $v = D + \mu$ in the opposite regime
$D/\mu \to 0$ (diffusion-limited regime). In Fig.\ \ref{vel} we show
the results of our simulations for the velocity of the front $v$ as a
function of $\mu$ for different values of $\eta$.  Our results are
consistent with the exact results \cite{doering1,avraham,bramson86}
and previous simulations of this model \cite{kerstein}. For an
intermediate case $0<\eta<D$ we observe that, for some values of
$\mu$, the velocity seems to approach the deterministic value $v_0$.
However, for small enough value of $\mu$ internal fluctuations seem to
dominate and the velocity deviates strongly from $v_0$.

\begin{figure}
\begin{center}
\includegraphics[width=3.0in,clip=]{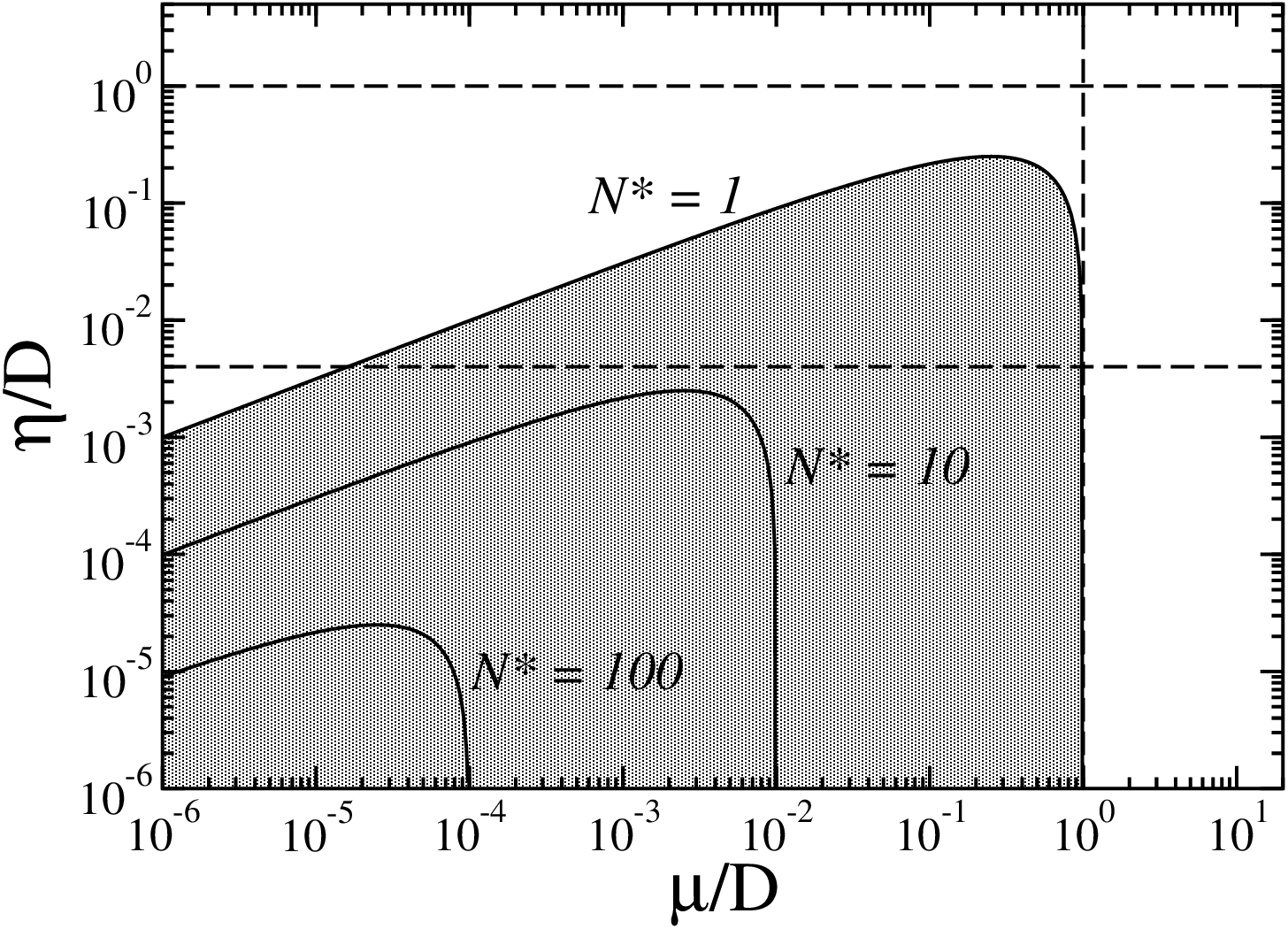}
\caption{\label{diag}
  Schematic diagram of condition (\ref{cond1}) in the $\eta,\mu$ phase
  space.  Shaded area correspond to those values of the parameters for
  which $N^* >1$. Solid lines correspond to the solution of Eq.\ 
  (\ref{cond2}) for different values of $N^*$. Dashed lines are the
  following sets of parameters: $\eta=D$ \cite{doering1,avraham1},
  $\mu=D$ \cite{panja} and $\eta = 0.002$ used in Fig.\ \ref{vel}.
}
\end{center}
\end{figure}

To understand this behavior, let us recall that mean field
approximation (\ref{FKPPeq}) in the $A \rl A + A$ is only valid when
diffusive mixing is strong enough.  Specifically, this happens when
the typical distance traveled diffusively by a particle between
reaction events, $l_D$, is much larger that the typical distance
between particles, $l_p$ \cite{avraham}. In that case, particles are
well stirred within cells of size $l_D$ and thus, mean-field
approximation is valid for the coarse-grained density of particles
over cells of size $l_D$ as shown in \cite{bramson86}. In our model we
have that $l_D = \min(\sqrt{D/\mu},\sqrt{D/\eta})$, while we
approximate $l_p$ by the average distance between particles in the
stable phase $l_p = \rho_0^{-1} = (\mu+\eta)/\mu$. Since we are
interested in the propagation of pulled fronts, which are only driven
by the birth term, our condition to approach the mean-field
approximation is then given only by $\mu$:
\begin{equation}\label{cond1}
\sqrt{D/\mu} \gg \rho_0^{-1} = \frac{\mu + \eta}{\mu}.
\end{equation}
Interestingly, this condition is equivalent to $N^* \equiv
\lambda^{-1} \rho_0 \gg 1$, where $N^*$ is approximately the number of
particles within an interval of length $\lambda^{-1}$.  Thus condition
(\ref{cond1}) also means that the number of particles within the
typical length scale of the front (its width) is large. Our results in
Fig.\ \ref{vel} are then easily explained in terms of condition
(\ref{cond1}): when $N^* \gg 1$ internal fluctuations should be
unimportant within cells of size $l_D \simeq \lambda^{-1}$ and front
propagation should approach asymptotically the FKPP predictions.
Since
\begin{equation}\label{cond2}
N^* = \frac{\sqrt{\mu/D}}{\mu/D + \eta/D},
\end{equation}
then $N^* \gg 1$ only happens when $(D/\eta)^2 \gg D/\mu \gg 1$ for
fixed values of $D$ and $\eta$. We show the condition $N^* > 1$ in
Fig.\ \ref{diag} along with the different set of parameters used in
Fig.\ \ref{vel} and in other works \cite{doering1,avraham1,panja}.
Outside the region $N^* > 1$ internal fluctuations dominate and fronts
are not described by the FKPP equation. This is the case for $\eta =
D$ \cite{doering1,avraham} and $\eta = \mu$ \cite{panja}. In the
intermediate case $0<\eta<D$ we can have values of $\mu$ for which
$N^*$ is relatively large and fronts seem to approach the
deterministic value of $v_0$, which explains the behavior observed in
Fig.\ \ref{vel} for $\eta = 0.002$.  Note however, that although being
in the region $N^* > 1$ is the minimum requirement for our model to
approach the mean-field description (\ref{FKPPeq}) a finite value of
$N^*$ means that fronts are still subject to internal fluctuations and
discreteness effects which produce a (strong) correction to the
velocity. Only in the limit $N^* \to \infty$ do both effects become
negligible and the $A\rl A+A$ system is effectively described by the
FKPP equation. This is the case for $\eta = 0$, $\mu \to 0$
\cite{note}.

\begin{figure}
\begin{center}
  \includegraphics[width=3.0in]{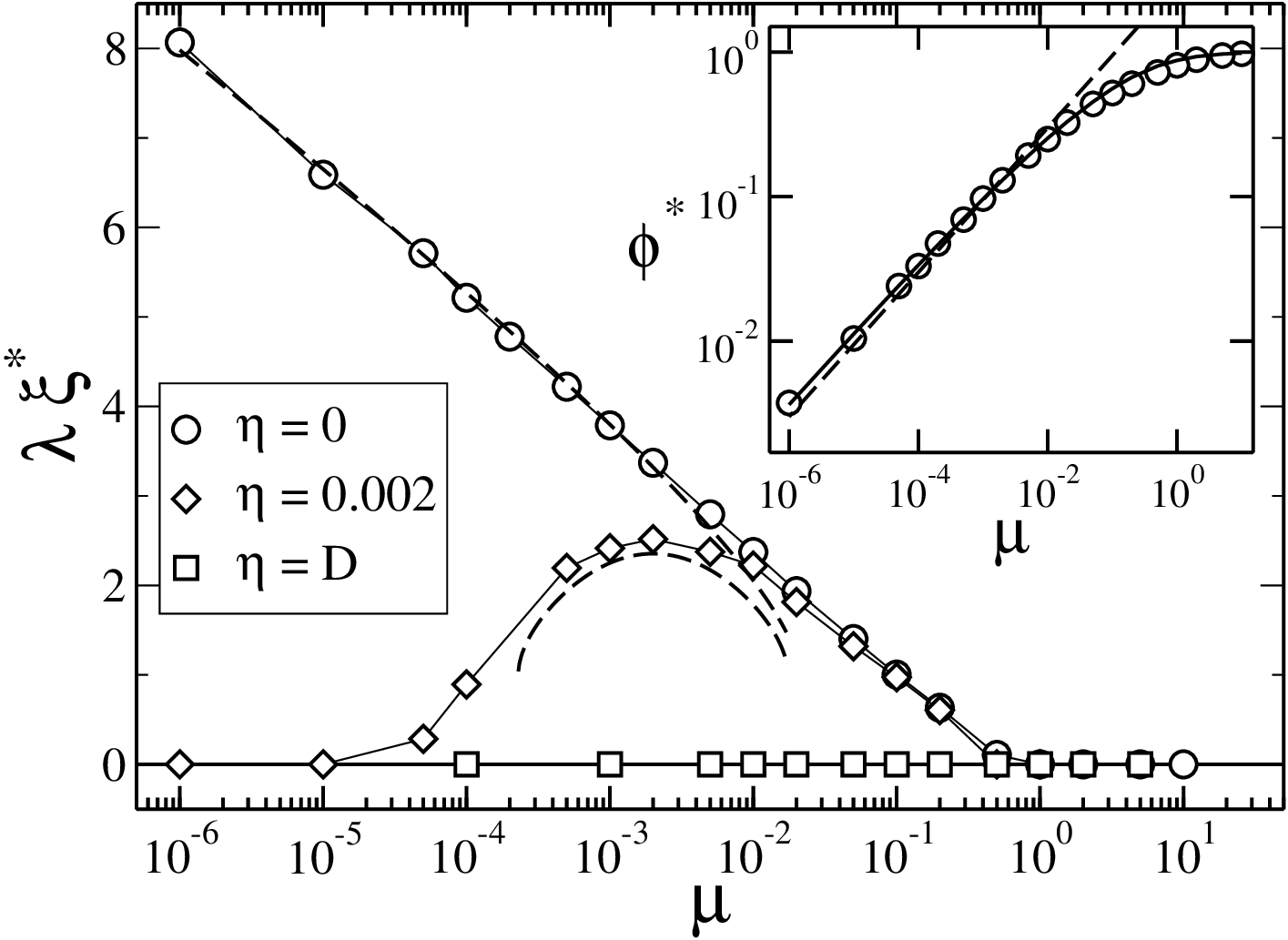}
\caption{\label{xi}
  Position of the last particle from the front position, $\xi^*$, as a
  function of $\mu$ from simulations. Dashed lines are Eq.\
  (\ref{xi_star}) with $a = 1.4$.  Inset: occupation probability for
  the site just before the last particle (circles). Solid line is
  given by Eq.\ (\ref{vel_chorra}) with $v$ taken from simulations and
  dashed line is the approximation $\rho^* \simeq a/\lambda$ used to
  get Eq.\ (\ref{xi_star}) \cite{note2}.}
\end{center}
\end{figure}

An interesting question is whether $N^*$ plays any role like the
average number of particles per site, $N$, in bosonic models
\cite{brunet,kessler,breuer}. In those models, it is observed that the
deterministic description of a pulled front given by the FKPP equation
is valid until the density drops to $\rho \simeq N^{-1}$ which
produces an effective cutoff in the tail of the front and modifies its
velocity \cite{brunet}.  To check this possibility, we have measured
in our simulations the average distance of the last particle from the
front position, $\xi^*$, which is observed to saturate to a constant
value for long enough times. It is obvious that for $\xi > \xi^*$ the
continuum description of the front breaks down and we expect this to
happen when there is only a particle in each coarse-grained site of
length $\lambda^{-1}$, i.e.\ when $\rho(\xi^*) \simeq a\lambda$, where
$a$ is a constant. When internal fluctuations are irrelevant, i.e.
when $N^* \gg 1$, we assume that the continuum description
(\ref{FKPPeq}) is still valid up to $\xi^*$ and taking that $\rho
\simeq \rho_0 \lambda \xi e^{-\lambda \xi}$ for $\lambda \xi \geq 1$
for a pulled front \cite{ebert} we obtain
\begin{equation}\label{xi_star}
\lambda \xi^* e^{-\lambda \xi^*} = a/N^*.
\end{equation}
Solutions of this equation for $\xi^*$ with $N^*$ given by
(\ref{cond2}) are compared with our simulations in Fig.\ \ref{xi}. We
see that for $\lambda \xi^* \gtrsim 1$ Eq.\ (\ref{xi_star}) gives a
rather accurate prediction of $\xi^*$. This corroborates our
assumption that a pulled front described by the FKPP equation
develops even for moderate values of $N^*$ up to the point where
$\rho \simeq (N^*)^{-1}$.


An important consequence of Eq.\ (\ref{xi_star}) is that $N^*$
controls not only the size of internal fluctuations but also the
appearance and length of the tail in the pulled front. Thus, when $N^*
\lesssim 1$ internal fluctuations dominate and also the tail length is
roughly zero, $\lambda \xi^* \simeq 0$. This means that the front is
basically a shock wave with height $\rho_0$.  Actually, the exact
solution when $\eta = D$ \cite{avraham1} shows that the particles
behind the leading one remain distributed as in the stable phase
($\rho = \rho_0$) at all times, which confirm our picture (see Fig.\ 
\ref{xi}). This {\em shock wave} shape is also observed in the case
$\eta \neq D$ when $\mu \gg \eta$, where it is found that $\lambda \xi
\simeq 0$. In the intermediate case in which $\eta \neq 0$ we see that
the front develops a tail which is described by the FKPP equation only
for a given interval of values of $\mu$ (see Fig.\ \ref{xi} for $\eta
= 0.002$).

In the case $\eta=0$ we have studied the correction to the velocity as
a function of $N^*$ and observe in the inset Fig.\ \ref{vel} that it
decays like $(v_0-v)/v_0 \sim (N^*)^{-1/3}$ which is consistent with
simulations of other microscopic bosonic models
\cite{lemarchand,brunet} for moderate values of number of particles
$N$.  This results stresses the equivalence of $N^*$ with the role
that the number of particles plays in other microscopic models. In
particular we expect the correction to be $(v-v_0)/v_0 \sim \ln^{-2}
N^*$ for very large values of $N^*$ \cite{note1}.

Since the last particle is, on average, at a certain distance from the
front position, their velocities coincide. This fact was used in
\cite{kerstein,panja} to estimate the velocity of the front by
counting possible forward and backward hopping rates:
\begin{equation}\label{vel_chorra}
v = \mu - \rho^* (\eta - D)
\end{equation}
where $\rho^*$ is the probability of having a particle behind the last
one. Several approximations can be made for the value of $\rho^*$
\cite{kerstein,panja}. For example, in \cite{panja} is was taken as
$\rho^* \simeq \rho_0$, i.e. $\rho^*$ is given by the probability to
find a particle in the stable phase.  Clearly, this approximation is
only valid in the case in which fronts are like a shock-wave, i.e.\ 
when $N^*\simeq 1$ because then the last particle is very close to the
stable phase. In the case in which a pulled front  develops ($N^*
\gg 1$) we find that Eq.\ (\ref{vel_chorra}) still holds: since the
last particle is on average at a fixed distance from the front
position, we can approximate $\rho^* \sim \rho_0 \lambda\xi^*
e^{-\lambda \xi^*} = a/\lambda$ which is the concentration of
particles at $\xi^*$. Our simulations for $\eta =0$ confirm the
validity of this approximation (see Fig.\ \ref{xi}) which brings out
the effective matching between the continuum description given by the
FKPP equation for $\xi < \xi^*$ and the microscopic character of
the model for $\xi \gtrsim \xi^*$.

\begin{figure}
\begin{center}
\includegraphics[width=3.2in,clip=]{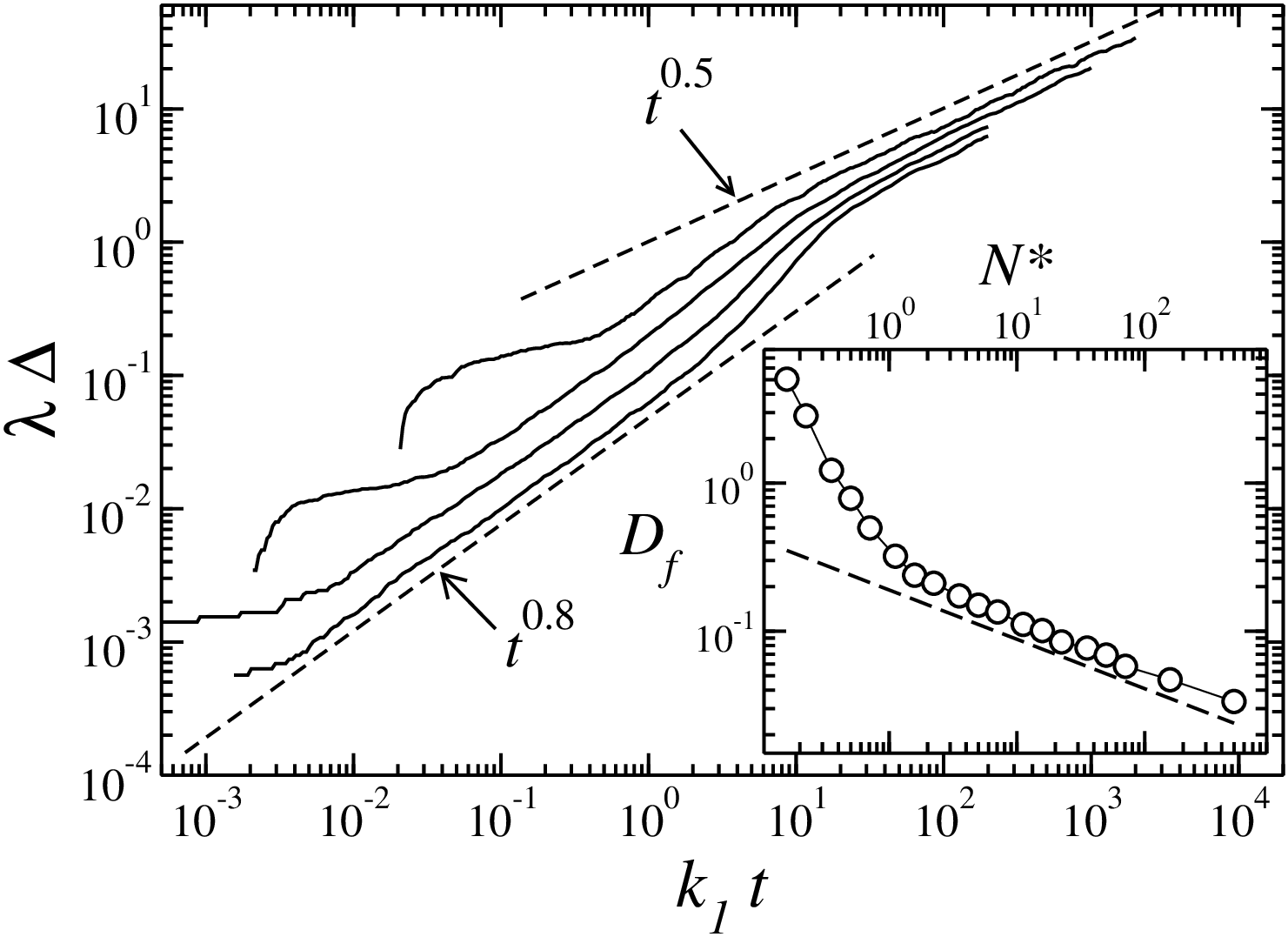}
\caption[phi2]{\label{delta}
  Time evolution of the diffusive spreading of the front position for
  different values of $\mu$ (solid lines) and $\eta =0$:
  $\mu=10^{-2},10^{-3},10^{-4},10^{-5}$ from top to bottom. Inset:
  Diffusion coefficient as a function of $N^*$ for $\eta=0$. Dashed
  line is the power law $(N^*)^{-1/3}$.}
\end{center}
\end{figure}

Another interesting property of front propagation is the wandering of
the position of the front around its mean value $\Delta^2(t) =
\mean{(x_f(t)-\mean{x_f(t)})^2}$. Several studies of different
microscopic models shows that $\Delta^2(t) = 2 D_f t$ for long times
\cite{breuer,brunet,lemarchand} and that the diffusion coefficient
$D_f$ depends on the number of particles $N$.  Specifically, it was
found that $D_f \sim N^{-1/3}$ for moderate values of $N$
\cite{breuer,lemarchand}, while $D_f \sim \ln^{-3} N$ for very large values
of $N$ \cite{brunet}. Our simulations for the $A+A\rl A$ system shows
that fronts move diffusively for all values of the parameters.  In the
case $\eta=0$, in which the model approaches asymptotically the FKPP
equation, we get $D_f \sim (N^*)^{-1/3}$ (see Fig.\ \ref{delta}) like
in bosonic models \cite{lemarchand} which stress once again the fact
that $N^*$ plays the role of the number of particles in this fermionic
model.  Moreover, we found for small times that as $N^*$ increases,
the correlation between the time development of the front and internal
fluctuations produces superdiffusive motion of the front position
$\Delta^2(t) \sim t^{2\nu}$ with $\nu \simeq 0.8$. Once the front tail
is developed (which happens at $t \simeq k_1^{-1}$), front position
starts to wander diffusively. Finally, our results for the diffusion
of the front indicate that as the front approaches the deterministic
FKPP equation, internal fluctuations make the front move diffusively
at times $t > k_1^{-1}$, independently of $N^*$. We do not observed
any signs of subdiffusive behavior conjectured by some authors
\cite{tripathy,saarloos} for pulled fronts subject to noise.  This
supports the idea that pulled fronts subject to internal noise belong
to a different universality class than those subject to external noise
\cite{moro}.

In summary, we have identified the parameter that controls the
strength of microscopic fluctuations for the front propagation problem
in the fermionic model $A \rl A+A$, namely the number of particles
$N^*$ per coarse-grained site of length $\lambda^{-1}$.  When $N^* \gg
1$, internal fluctuations are suppressed and the front becomes a
pulled front like those of the FKPP equation (\ref{FKPPeq}). Moreover,
our studies about the length of the tail, the velocity of the front
and its diffusion show that $N^*$ plays the same role as the number of
particles in other microscopic bosonic models. Finally, it is
interesting to note that in the $A \rl A+A$ model, the velocity of a
macroscopic object such as the front is related to the microscopic
motion of the last particle, something also observed in other works
\cite{brunet,panja,kerstein88}. We hope our results will help to
understand the dynamics of fronts in microscopic fermionic
reaction-diffusion models and its relevance when discussing properties
of the FKPP equation subject to internal noise
\cite{tripathy,saarloos,moro}.

  We are grateful to D.\ ben-Avraham, E.\ Brunet, R.\ Cuerno, J.\
  Casademunt, C.\ Doering and W.\ van Saarloos for comments and
  discussions. Financial support is acknowledged from the European
  Commission through its Marie Curie program and from the Ministerio de
  Ciencia y Tecnolog\'\i a (Spain).


\begin{thebibliography}{99}
\bibitem{note} Note however that other scalings like $\eta \sim \mu$
  in Eq.\ (\ref{cond2}) are possible to get $N^*\to \infty$ in the
  limit $\mu \to 0$.
\bibitem{note1} A crude estimation for this to happen is when
  $N^{-1/3} \simeq \pi^2/(2 \ln^2 N^*)$ \cite{brunet} which gives $N^*
  \sim 10^4$, beyond our range of values for $N^*$.
\bibitem{note2} Due to the coarse-graining over intervals of length
  $\lambda^{-1}$ to calculate the front position, $\xi^*$ could be
  negative. In that case we take $\xi^* = 0$ since it is obvious that
  the last particle is the one that defines the front position.
\end{thebibliography}
\end{document}